\documentclass[12pt,preprint]{aastex}
\usepackage{epsfig}
\begin{document}

\title{Excitation and Propagation of Eccentricity Disturbances in Planetary Systems}

\author{
Nadia L. Zakamska\altaffilmark{1}, 
Scott Tremaine\altaffilmark{1}}
\altaffiltext{1}{Princeton University Observatory, Princeton, New Jersey 08544}

\begin{abstract}
The high eccentricities of the known extrasolar planets remain largely unexplained. We explore the possibility that eccentricities are excited in the outer parts of an extended planetary disk by encounters with stars passing at a few hundreds of AU. After the encounter, eccentricity disturbances propagate inward due to secular interactions in the disks, eventually exciting the innermost planets. We study how the inward propagation of eccentricity in planetary disks depends on the number and masses of the planets and spacing between them and on the overall surface-density distribution in the disk. The main governing factors are the large-scale surface-density distribution and the total size of the system. If the smeared-out surface density is approximated by a power-law $\Sigma(r)\propto r^{-q}$, then eccentricity disturbances propagate inward efficiently for flat density distributions with $q\la 1$. If this condition is satisfied and the size of the planetary system is 50 AU or larger, the typical eccentricities excited by this mechanism by field star encounters in the solar neighborhood over 5 Gyr are in the range 0.01-0.1. Higher eccentricities ($>$0.1) may be excited in planetary systems around stars that are formed in relatively dense, long-lived open clusters. Therefore, this mechanism may provide a natural way to excite the eccentricities of extrasolar planets.
\end{abstract}


\section{Introduction}

One of the remarkable features of the $\sim 120$ known extrasolar planets is their relatively high eccentricities\footnote{A catalog of extrasolar planets is maintained by Paris Observatory (http://www.obspm.fr/planets)}, most far larger than seen in the giant planets of the solar system. Many mechanisms to produce these high-eccentricity orbits have been offered. The most popular involve planet-planet gravitational scattering \citep{rasi96,weid96,lin97,ford01,yu01,chia02, terq02}, either through close encounters, physical collisions, secular interactions or resonant interactions. Other eccentricity excitation mechanisms include interactions with the gaseous protoplanetary disk \citep{gold80,papa01,gold03,ogil03}, or interactions with the planetesimal disk \citep{murr98}.

All of these proposed mechanisms have shortcomings. Planet-planet scattering through close encounters appears to create the wrong eccentricity distribution---there may be too many planets on low-eccentricity orbits, formed by physical collisions or tidal captures during the scattering process (\citealt{ford01,gold03}, but see \citealt{ford03}). Discussions of eccentricity excitation through mean-motion resonances assume that planet-disk interactions induce evolution of semimajor axes leading to resonant capture, but neglect the effects of these interactions on eccentricity. Interactions with the gaseous planetesimal disk can either damp or excite eccentricity, depending sensitively on the formation of gaps at Lindblad resonances, nonlinear saturation of corotation resonances \citep{ogil03}, and the rate of viscous diffusion across resonances \citep{gold03}, all of which are difficult to calculate reliably. Excitation through interactions with the planetesimal disk may require more massive disks than are indicated by observations and may only be efficient for companions with masses much higher than those of most extrasolar planets \citep{papa01}.

There is strong and growing evidence that most forming stars are surrounded by extended disks, with radii of hundreds of AU. This evidence includes CO observations of massive gas disks around young stars, disks detected in Orion and other star-forming regions by optical emission from their photoionized surfaces, the infrared excess at wavelengths of 25--100$\mu$ discovered by IRAS around bright nearby stars such as Vega, dust disks found optically around nearby stars such as $\beta$ Pictoris, and imaging by the NICMOS camera on the Hubble Space Telescope of near-infrared radiation scattered by dusty disks around stars such as HR 4796 (see \citealt{koer01} and \citealt{zuck01} for reviews). It may well be that the dust around older stars such as Vega and $\beta$ Pictoris is ``second-generation'' dust, released in collisions between larger objects (planetesimals or comets) that were formed by the accretion of primordial or first-generation dust particles. Thus the absence of dust around older stars may arise either because the dust has been dispersed or destroyed or because it has been incorporated into larger bodies such as planets. 

Indirect evidence for planets at large distances from the host stars comes from the morphologies of the dust debris disks of Vega and $\varepsilon$ Eri. Disk asymmetries and dust concentrations are often interpreted as resulting from the dynamical influence of an unseen massive planet. Modeling suggests that the planetary companion has a semimajor axis $a>30$ AU in the Vega system \citep{wiln02} and about 55--65 AU in the $\varepsilon$ Eri system \citep{ozer00}. Formation of giant planets by core accretion is difficult at these distances, and it has been suggested that planets can migrate outward to distances of up to tens of AU through interactions with the gas disk in young systems \citep{vera04}.

We propose to explore the possibility that the eccentricities of the extrasolar planets arise because of secular interactions in a long-lived extended planetary or protoplanetary disk. We suggest that eccentricities are excited in the outer part of the disk by a passing star and propagate inward through the disk, somewhat like a wave. We explore how the inward propagation of eccentricity in disks depends on the overall surface-density distribution.

In Section \ref{sec_secular} we review the definitions and the approach of secular perturbation theory. In Section \ref{sec_excit} we describe the initial excitation of the eccentricity and in Section \ref{sec_prop} we study the propagation of eccentricities via secular interactions. We discuss typical eccentricities produced by this mechanism in Section \ref{sec_disc} and conclude in Section \ref{sec_conc}.

\section{Secular evolution of a planetary system}
\label{sec_secular}

In this Section we consider a planetary system around a star with mass $M_c$. There are $N$ planets with masses $m_i\ll M_c$ and orbital elements $a_i$ (semimajor axes), $\varepsilon_i$ (eccentricities), $\varpi_i$ (longitudes of pericenter) and $n_i$ (mean motions), where $n_i^2 a_i^3=G(M_c+m_i)$. The index $i$ runs from the inner to the outer planet, and sets of $N$ values can be treated as vectors. We assume that the parameters of the planetary system are such that the mean-motion resonances can be neglected. So long as the eccentricities and inclinations of the orbits remain small, the secular evolution of the system can be investigated using Laplace-Lagrange theory \citep{brow61, murr99}. Specifically, with the change of variables 
\begin{equation}
h_i=\varepsilon_i\sin\varpi_i, \quad k_i=\varepsilon_i\cos\varpi_i\label{change}
\end{equation}
the equations that describe the evolution of eccentricities and the motions of the pericenters of the orbits are 
\begin{equation}
\dot{h}_i=\sum_{j=1}^N {\mathbb A}_{ij} k_j;\quad \dot{k}_i=-\sum_{j=1}^N {\mathbb A}_{ij} h_j.\label{dynam}
\end{equation}
The matrix $\mathbb A$ is completely determined by the masses and semimajor axes of the planets and the mass of the central star. In this approximation, eccentricities $\varepsilon_i$ and longitudes of pericenter $\varpi_i$ are not coupled to inclinations $I_i$ and longitudes of ascending nodes $\Omega_i$. The planets can be treated as coupled harmonic oscillators, and the time evolution of $h_i$ and $k_i$ is a superposition of oscillations with frequencies determined by the eigenvalues of the matrix ${\mathbb A}$ and with amplitudes determined by the initial conditions and the eigenvectors of ${\mathbb A}$. Equations similar to (\ref{dynam}) describe evolution of inclinations and longitudes of the ascending nodes. 

More specifically, the matrix ${\mathbb A}$ is \citep{brow61}
\begin{eqnarray}
{\mathbb A}_{ij}=-\frac{1}{4} n_i \frac{m_j}{M_c+m_i} \alpha_{ij} \tilde{\alpha}_{ij} b_{3/2}^{(2)}(\alpha_{ij})\quad (i\ne j),\nonumber\\
{\mathbb A}_{ii}=\frac{1}{4} n_i \sum_{j=1,j\ne i}^N \frac{m_j}{M_c+m_i} \alpha_{ij} \tilde{\alpha}_{ij} b_{3/2}^{(1)}(\alpha_{ij}), \label{A}
\end{eqnarray}
where
\begin{equation}
\alpha_{ij}=\min \left(\frac{a_i}{a_j},\frac{a_j}{a_i}\right),\quad 
\tilde{\alpha}_{ij}=\min \left(\frac{a_i}{a_j},1\right)
\end{equation}
and
\begin{equation}
b_s^{(\nu)}(\alpha)=\frac{1}{\pi}\int_0^{2 \pi}\frac{\cos(\nu x)dx}{(1-2\alpha\cos x+\alpha^2)^s},\quad \alpha<1.
\end{equation}
are Laplace coefficients. With $\omega_{\mu}$ being the eigenvalues of ${\mathbb A}$ and ${\bf e}_{\mu}$, $\mu=1...N$ being its eigenvectors, the solutions of equations (\ref{dynam}) are 
\begin{eqnarray}
h_i(t)=\sum_{\mu=1}^N e_{\mu, i} S_{\mu} \sin [\omega_{\mu} (t-t_0) + \beta_{\mu}];\label{sol1}\\
k_i(t)=\sum_{\mu=1}^N e_{\mu, i} S_{\mu} \cos [\omega_{\mu} (t-t_0) + \beta_{\mu}];\label{sol2}\\
\varepsilon_i(t)=\sqrt{h_i^2(t)+k_i^2(t)}.\label{sol3}
\end{eqnarray}
The amplitude coefficients $S_{\mu}$ and the phases $\beta_{\mu}$ are determined by the initial conditions $h_i(t_0)$ and $k_i(t_0)$. 

The initial conditions and the eigenvectors can be weighted using $w_i=a_i\sqrt{m_i n_i}$:
\begin{equation}
H_i(t_0)=h_i(t_0) w_i;\quad K_i(t_0)=k_i(t_0) w_i; \quad E_{\mu,i}=e_{\mu,i} w_i.\label{weight}
\end{equation} 
It can be shown that the set of vectors ${\bf E}_{\mu}$ is orthogonal and that the weighted eccentricity components $H_i$ and $K_i$ satisfy equations similar to (\ref{dynam}) with an orthogonal matrix $\mathbb A'$.

The mean square eccentricity $\left<\varepsilon^2_i(t)\right>=\left<h_i^2(t)+k_i^2(t)\right>$ of each planet provides a time-independent measure of the level of excitation. This value can be computed from the eigenvectors of the matrix $\mathbb A$ and the eccentricity components for all planets at some time $t_0$:
\begin{equation}
\left<\varepsilon_i^2(t)\right>=\sum_{\mu=1}^N \left(e_{\mu,i}\right)^2 \frac{({\bf H}(t_0)\cdot{\bf E}_{\mu})^2+({\bf K}(t_0)\cdot{\bf E}_{\mu})^2 }{({\bf E}_{\mu}\cdot{\bf E}_{\mu})^2}.\label{eccent}
\end{equation}

Mean square eccentricities can be used to quantify the response of the system to an external perturbation -- for example, another star passing not far from the planetary system. Before and after the perturbation, when the perturber is far enough so that its potential can be neglected compared to the potential of the central star and the planets, the planetary system evolves according to secular perturbation theory, as long as eccentricities and inclinations remain small. In this case, the mean square eccentricities after the perturbation are
\begin{equation}
\left<\varepsilon_i^2(t)\right>_{after}=\sum_{\mu} \left(e_{\mu,i}\right)^2 \frac{\left(({\bf H}(t_0)+\delta {\bf H}) \cdot{\bf E}_{\mu}\right)^2+\left(({\bf K}(t_0)+\delta {\bf K})\cdot{\bf E}_{\mu}\right)^2 }{({\bf E}_{\mu}\cdot{\bf E}_{\mu})^2}.\label{eccent_p}
\end{equation}
Here $\delta h_i$ and $\delta k_i$ are the changes in the eccentricity components due to the perturbation that are weighted similarly to (\ref{weight}) to obtain $\delta H_i$ and $\delta K_i$. 

Below we focus on excitation of eccentricities of planets on initially circular orbits, so that ${\bf H}(t_0)=0$ and ${\bf K}(t_0)=0$.

\section{Excitation of eccentricities by an external perturber}
\label{sec_excit}

In this Section we calculate the response of the planets to an external perturber, which is quantified by the values $\delta h_i$ and $\delta k_i$ in the notation of the previous Section. This response can be obtained by numerical integration of the equations of motion in an arbitrary geometry of the perturbation but for simplicity we consider the two-dimensional case, when the perturber, the central star and the planets are coplanar at all times. We also neglect the interaction between the planets during the perturbation. This is justified as long as the characteristic timescale of the perturbation is much shorter than the timescales of secular evolution, which we estimate as inverse secular eigenfrequencies of the system $P_{secular}\sim P_{orbital, i}M_c/m_i$. 

\begin{figure}
\epsscale{0.7}
\plotone{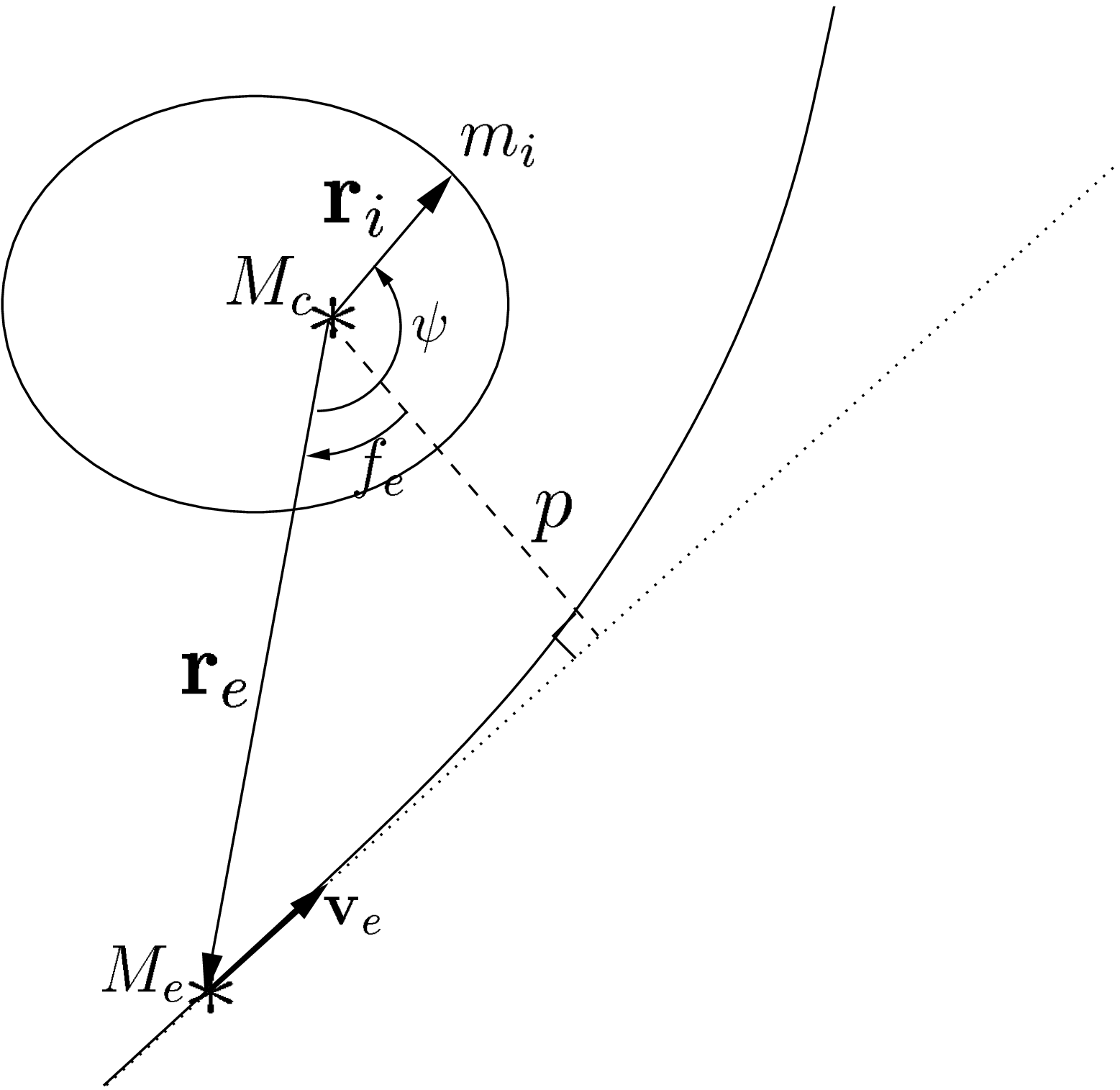}
\figcaption{Geometry of the encounter in the reference frame of the host star $M_c$: the planetary system orbits the star $M_c$ and the passing star $M_e$ is a perturber moving on a hyperbolic trajectory. ${\bf v}_e$ is the velocity of the perturber far from $M_c$, and $p$ is its impact parameter. In this Figure, the motion of the planet relative to the perturber is prograde if it is moving counter-clockwise and retrograde otherwise.\label{pic_impact}}
\end{figure}

The perturber has a mass $M_e$ and when it is far from the system it moves with velocity $v_e$ and impact parameter $p$ relative to the central star (Figure \ref{pic_impact}). The equations of motion of the external perturber and of the planet $m_i$ relative to the central star are
\begin{eqnarray}
\ddot{\bf r}_e=-G\frac{(M_c+M_e){\bf r}_e}{r_e^3};\nonumber \\
\ddot{\bf r}_i=-G\frac{M_c {\bf r}_i}{r_i^3}-G\frac{M_e ({\bf r}_i-{\bf r}_e)}{|{\bf r}_i-{\bf r}_e|^3}-G\frac{M_e {\bf r}_e}{r_e^3}.\label{impact}
\end{eqnarray}
Here the contribution to the potential from the planets is neglected during the perturbation; the planets move as test particles in the potential field of the central star and the external perturber. The trajectory of the perturber relative to the central star is a hyperbola. We restrict ourselves to the cases when the perturber passes outside the planetary system, so that the distance of the closest approach of the perturber is $>a_N$.

The excited eccentricities remain the same if the parameters are rescaled so that $a\rightarrow \kappa a$, $p\rightarrow \kappa p$, $v_e \rightarrow \kappa^{-1/2} v_e$ because the timescale of the perturbation $\tau_e=p/v_e$ scales as $\tau_e\rightarrow \kappa^{3/2}\tau_e$, in the same way as does the orbital period of the planet. This can be seen from equations (\ref{impact}) which are invariant to such change of variables. 

As a result of the interaction with the central star the perturber deviates from its original trajectory by an angle
\begin{equation}
\chi=\pi-2 \arctan \frac{v_e^2 p}{G(M_c+M_e)}.\label{chi}
\end{equation}
A perturbing star that moves with a velocity
\begin{equation}
v_e\gg\sqrt{\frac{G (M_c+M_e)}{p}} \label{cond_fast}
\end{equation}
proceeds almost along a straight line with almost constant velocity. In this case (which we refer to as a ``fast perturber'') some simple approximate solutions of equations (\ref{impact}) can be obtained analytically. In Section \ref{fast_perturber} we present analytical estimates and numerical solutions for the case of the fast perturber and in Section \ref{slow_perturber} discuss the general case when the perturber does not satisfy condition (\ref{cond_fast}). 

\subsection{Fast perturber}
\label{fast_perturber}

In the solar neighborhood, the velocity dispersion $\sigma\simeq 40$ km sec$^{-1}$, and given the stellar density of about $n_*=0.05$ pc$^{-3}$, about half of all solar-type stars have experienced an encounter within 500 AU at least once during the last 5 Gyr (the Sun's current age). The average encounter velocity is $4 \sigma/\sqrt{\pi}$ in the approximation that the distribution function is isotropic and Maxwellian. All these encounters are fast by our definition (\ref{cond_fast}), since in this case $\chi$ (eq. \ref{chi}) is only a few $\times 10^{-3}$ rad. 

{\bf Impulse approximation}

If the perturber is fast and if in addition the orbital period of the planet is much longer than the characteristic timescale of the perturbation $\tau_e=p/v_e$, then the displacement of the planet and the force from the central star can be neglected during the perturbation (``impulse approximation''). Relative to the central star, the planet effectively receives a kick in velocity directed perpendicular to the trajectory of the perturber:
\begin{equation}
\Delta v_{\perp, i}=\frac{2 G M_e}{v_e p}\frac{a_i\cos\varphi_i}{p-a_i\cos\varphi_i},
\end{equation}
where $\varphi_i$ is the angle between ${\bf r}_i$ and ${\bf r}_e$ at the instant of closest approach of the perturber to the central star. If the planet was initially on a circular orbit, then after the impact it proceeds on an elliptical orbit with eccentricity
\begin{equation}
\varepsilon_i=\frac{2 G M_e}{v_e p}\sqrt{\frac{a_i}{G M_c}}\left[\frac{a_i}{p}\sqrt{\cos^2\varphi_i(1+3\sin^2\varphi_i)}+O\left(\frac{a_i^2}{p^2}\right)\right].\label{fast}
\end{equation}
For $p\gg a_i$ this value reaches its maximum at $\varphi_{max}$ such that $\cos(2\varphi_{max})=1/3$ and its minimum at $\varphi_{min}=\pm\pi/2$. In the impulse approximation, the excited eccentricity does not depend on the direction of motion of the planet relative to the perturber (prograde, retrograde) -- the effect of the direction is only important when the orbital motion of the planet during the perturbation is significant. 

{\bf Secular approximation}

In the opposite limit the orbital period of the planet is short compared to the characteristic timescale of the perturbation $\tau_e=p/v_e$ (``secular approximation''). In this case the perturbation can be averaged over the orbital period, and thus we expect that the excited eccentricity is independent of the direction of motion of the planet relative to the perturber (prograde or retrograde) or the orbital phase of the planet at the moment of closest approach. The part of the disturbing function of the planet $i$ due to the external perturber is (following \citealt{murr99})
\begin{equation}
R_{i,ext}=\frac{GM_e}{|{\bf r}_e-{\bf r}_i|}-GM_e\frac{{\bf r}_i\cdot{\bf r}_e}{r_e^3}. \label{dist_function}
\end{equation}
We expand this expression in powers of $a_i/p$:
\begin{equation}
R_{i,ext}=\frac{GM_e}{r_e}\sum_{l=2}^{\infty}\left(\frac{r_i}{r_e}\right)^l P_l(\cos\psi),\label{radial}
\end{equation}
where $\psi$ is the angle between the vectors ${\bf r}_i$ and ${\bf r}_e$. For simplicity we only consider fast perturbers, so that they are not significantly deflected by the interaction with the central star. We take the first terms ($l=2,3$) of the expansion (\ref{radial}) and compute the time average of the disturbing function over one orbit of the planet:
\begin{eqnarray}
\left<R_{i,l=2}\right>=\frac{GM_e}{p}\left(\frac{a_i}{p}\right)^2\cos^3 f_e \left[\frac{1}{4}+\frac{3}{8}\varepsilon_i^2+\frac{15}{8}\varepsilon_i^2\cos(2\varpi_i-2f_e)+O(\varepsilon_i^3)\right];\label{l2}\\
\left<R_{i,l=3}\right>=\frac{GM_e}{p}\left(\frac{a_i}{p}\right)^3\cos^4 f_e \left[-\frac{15}{16}\varepsilon_i\cos(\varpi_i-f_e)+O(\varepsilon_i^2)\right].\label{l3}
\end{eqnarray} 
Here $f_e(t)$ describes the motion of the perturber and is a slow (compared to the orbital motion of the planet) function of time. For a perturber that moves almost along a straight line $f_e$ is given by 
\begin{equation}
\sin f_e(t)=\frac{v_e(t-t_a)}{\sqrt{p^2+v_e^2(t-t_a)^2}},
\end{equation}
$t_a$ being the time of the closest approach. 

We substitute expressions (\ref{l2})-(\ref{l3}) into the equations for the time evolution of eccentricities and positions of pericenters:
\begin{equation}
\dot{h}_i=+\frac{1}{n_ia_i^2}\frac{\partial \left<R_i\right>}{\partial k_i},\quad
\dot{k}_i=-\frac{1}{n_ia_i^2}\frac{\partial \left<R_i\right>}{\partial h_i}.
\end{equation}
Thus, in the presence of a coplanar external perturber the eccentricity components evolve according to
\begin{eqnarray}
\dot{h}_i\simeq\frac{GM_e}{p^3 n_i}\cdot \cos^3 f_e \left[k_i\left(\frac{15}{4}\cos 2f_e+\frac{3}{4}\right)+\frac{15}{4}h_i\sin 2f_e -\frac{15}{16}\frac{a_i}{p}\cos^2 f_e \right];\label{hdot}\\
\dot{k}_i\simeq- \frac{GM_e}{p^3 n_i}\cdot \cos^3 f_e \left[h_i\left(-\frac{15}{4}\cos 2f_e+\frac{3}{4}\right)+\frac{15}{4}k_i\sin 2f_e -\frac{15}{16}\frac{a_i}{p}\cos f_e \sin f_e\right],\label{kdot}
\end{eqnarray}
where again we have neglected interplanetary interactions during the perturbation. These are approximate equations in which terms on the order of
\begin{equation}
\frac{GM_e}{p^3 n_i}\cdot O\left(\varepsilon^2,\frac{a^2}{p^2},\varepsilon\frac{a}{p}\right)
\end{equation}
are neglected.

For a planet on an initially circular orbit, in equations (\ref{hdot})-(\ref{kdot}) the terms proportional to $k$ and $h$ can be neglected as long as $k,h\ll a/p$ during the event. In this case equations (\ref{hdot})-(\ref{kdot}) can be integrated to obtain
\begin{equation}
\delta h_i =-\frac{5}{4}\frac{GM_e}{p v_e}\sqrt{\frac{a_i}{GM_c}}\frac{a_i^2}{p^2}, \quad \delta k_i=0, \quad \delta\varepsilon_i = |\delta h_i|.\label{secular}
\end{equation}
This result can also be obtained from general formulas presented by \citet{hegg96} and \citet{koba01}\footnote{Due to a typographic error there is a factor of $e_*^{-3}$ missing in equations (46) and (47) of \citet{koba01}.} in the limit when the perturber is coplanar and moves very fast. 

\newpage
{\bf Intermediate regime}

If a planetary system has a large range of semimajor axes, it is plausible that the same passing star acts as an ``impulse'' perturber for the outer planets and as a ``secular'' perturber for the inner planets. Below we consider a case of a perturber with $M_e=M_{\odot}$ that has impact parameter $p=500$ AU and velocity $v_e=80$ km sec$^{-1}$. In this case the deflection angle of the perturber $\chi=1\times 10^{-3}$. We numerically integrated the equations of motion (\ref{impact}) for $M_c=M_{\odot}$ and obtained values of the excited eccentricities for a large range of semimajor axes (from 1 AU to 300 AU) and for several values of the orbital phase at closest approach $\varphi$ (cf.~\citealt{koba01}). The results are presented in Figure \ref{pic_excit} (top) together with the analytical approximations (\ref{secular}) and (\ref{fast}). 

The secular approximation (\ref{secular}) is shown with a dashed line. It is applicable for planets with orbital periods $P_{orbital}\ll \tau_e$, or in terms of semimajor axes $a\ll 7$ AU for the encounter with $p=500$ AU and $v_e=80$ km sec$^{-1}$, and the analytical formula (\ref{secular}) describes the excitation very well throughout its region of applicability. In the intermediate regime ($a$ of about tens of AU) the excited eccentricity depends both on the orbital phase at the closest approach of the perturber $\varphi$ and on the direction of the motion of the planet relative to the perturber, and the average excited eccentricity is substantially larger for prograde planets. For the slowest planets (large $a$) the impulse approximation (\ref{fast}) works reasonably well and the excited eccentricity only depends on the phase, not on the direction, in agreement with equation (\ref{fast}) and related discussion. The discrepancy between the theoretical prediction and the numerical calculation is due to neglecting terms of the next order in $a/p$ in (\ref{fast}) which become significant at $a\sim 100$ AU.

\subsection{Slow perturber}
\label{slow_perturber} 

Many or most stars are born in clusters. Open clusters are relatively short-lived ($\sim 10^8$ years) systems with velocity dispersions of $1-3$ km sec$^{-1}$ that reflect their small mass and with central densities about
\begin{equation}
n_*=80 \mbox{ pc}^{-3} \left(\frac{N}{1000}\right)\left(\frac{r_c}{1\mbox{ pc}}\right)^{-3},
\end{equation}
where $r_c$ is the cluster core radius \citep{king62, lyng82}. In such environments about half of the stars experience an encounter with impact parameter $p<400$ AU once during the lifetime of the cluster, but condition (\ref{cond_fast}) no longer holds, and deflection of the perturber becomes important. 

\begin{figure}
\epsscale{0.7}
\plotone{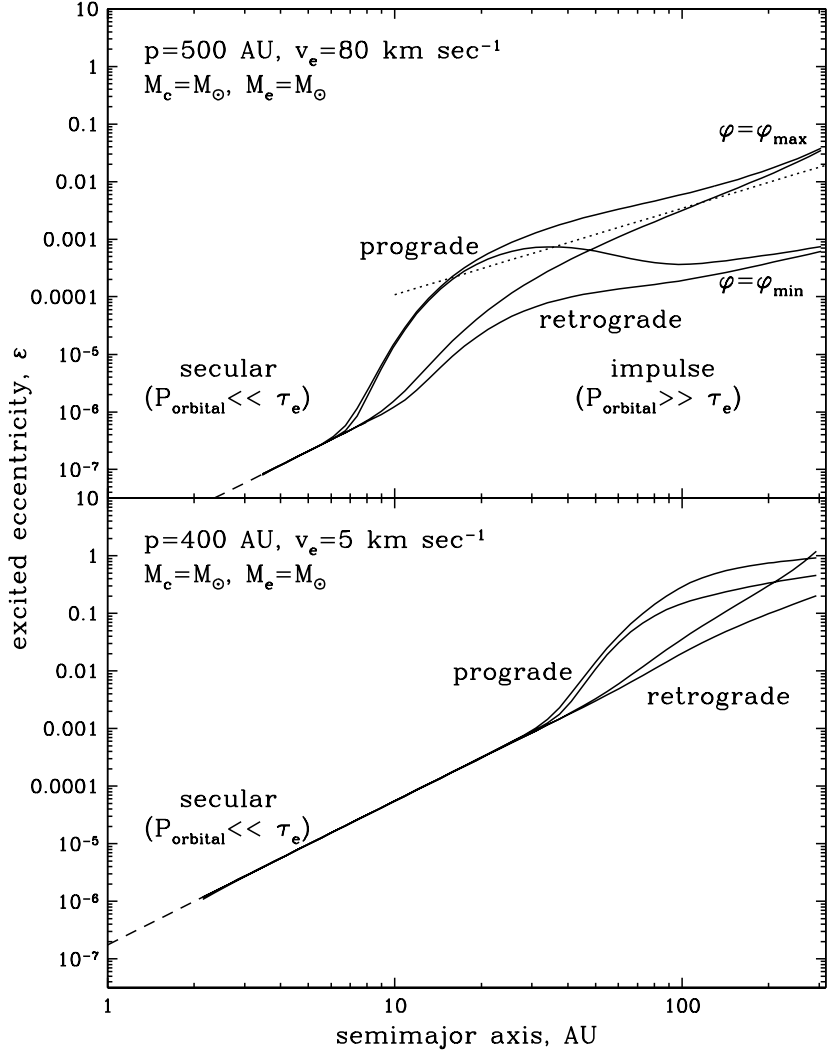}
\figcaption{Top: Excited eccentricity as a function of the semimajor axis $a$ of the planet for a fast encounter (with parameters typical for the solar neighborhood). Solid lines are results of the direct integration of the equations of motion for two values of the orbital phase $\varphi$ at the moment of closest approach of the perturber, and for two directions of the motion of the planet relative to the motion of the perturber (prograde and retrograde). Two extreme values of the phase (defined after eq. \ref{fast}) are chosen, so that the results for other phases lie in between the uppermost and the lowermost curves. The dashed lines show the analytical approximations in the secular regime (eq. \ref{secular}). The dotted line is the impulse approximation (\ref{fast}) plotted for $\varphi=\varphi_{max}$; it represents the theoretical maximum of the excited eccentricity. The results of the numerical integration for $\varphi=\varphi_{max}$ deviate from the dotted line because the assumption $a\ll p$ breaks down for large $a$. 
\newline Bottom: Same as above but for a slow encounter with parameters typical of rich open clusters. The dashed line is the secular approximation by \citet{hegg96} and \citet{koba01}. \label{pic_excit}}
\end{figure}

In Figure \ref{pic_excit} (bottom) we show the excited eccentricities calculated numerically from equations (\ref{impact}) as a function of semimajor axis in an impact with $p=400$ AU and $v_e=5$ km sec$^{-1}$. In this case the periastron distance of the perturber is 335 AU. Our secular approximation (\ref{secular}) is no longer valid because the perturber does not move along a straight line anymore, but generalized analytical solutions in the secular approximation were obtained by \citet{hegg96} and \citet{koba01}\footnote{Due to a typographic error there is a factor of $e_*^{-3}$ missing in equations (46) and (47) of \citet{koba01}.}:
\begin{equation}
\varepsilon_i=\frac{15}{8}\frac{M_e/M_c}{\sqrt{1+M_e/M_c}}\left(\frac{a_i}{p}\right)^{5/2}\frac{1}{\varepsilon_H (\varepsilon_H^2-1)^{5/4}}\left[\varepsilon_H^2 \arccos \left(-\frac{1}{\varepsilon_H}\right)+\frac{\sqrt{\varepsilon_H^2-1}}{3}(1+2\varepsilon_H^2)\right], \label{approx_rasio}
\end{equation}
where
\begin{equation}
\varepsilon_H=\sqrt{1+\left(\frac{p v_e^2}{G(M_e+M_c)}\right)^2}
\end{equation}
is the eccentricity of the hyperbolic motion of the perturber. Approximation (\ref{approx_rasio}) is plotted with a dashed line in Figure \ref{pic_excit} (bottom). 

Typical eccentricities excited during a slow interaction are one or two orders of magnitude higher than those excited by a fast perturber passing at similar distances. The expected number of encounters within a given impact parameter is similar in the solar neighborhood and in a rich open cluster; the large density of stars in the latter compensates for the short life-time and slow velocities. The result is that eccentricities acquired at the birthplace of stars during the first 10$^8$ years are much higher than those that can be generated during the subsequent evolution in a low-density environment. In what follows we will thus concentrate on slow encounters.

\section{Propagation of the disturbance via secular interactions}
\label{sec_prop}

After the perturber is gone, the excited eccentricities are redistributed in the system due to the secular interactions described in Section \ref{sec_secular}. As a first example, we consider a planetary system around a solar mass star consisting of 10 planets with their semimajor axes following a geometric progression from 1 AU to 100 AU, with $f=a_{i+1}/a_i=1.67$. The masses increase as a geometric progression, too, with the innermost planet having the mass of 0.1 $M_{Jupiter}$ and with $g=m_{i+1}/m_i=1.67$. The outer planet has then a mass of $m_N=10 M_{Jupiter}$. The planets were on circular orbits before the perturbation, and the orbital phases at the moment of the closest approach $\varphi_i$ were selected randomly. The perturber is in prograde motion relative to the planets. 

\begin{figure}
\epsscale{0.8}
\plotone{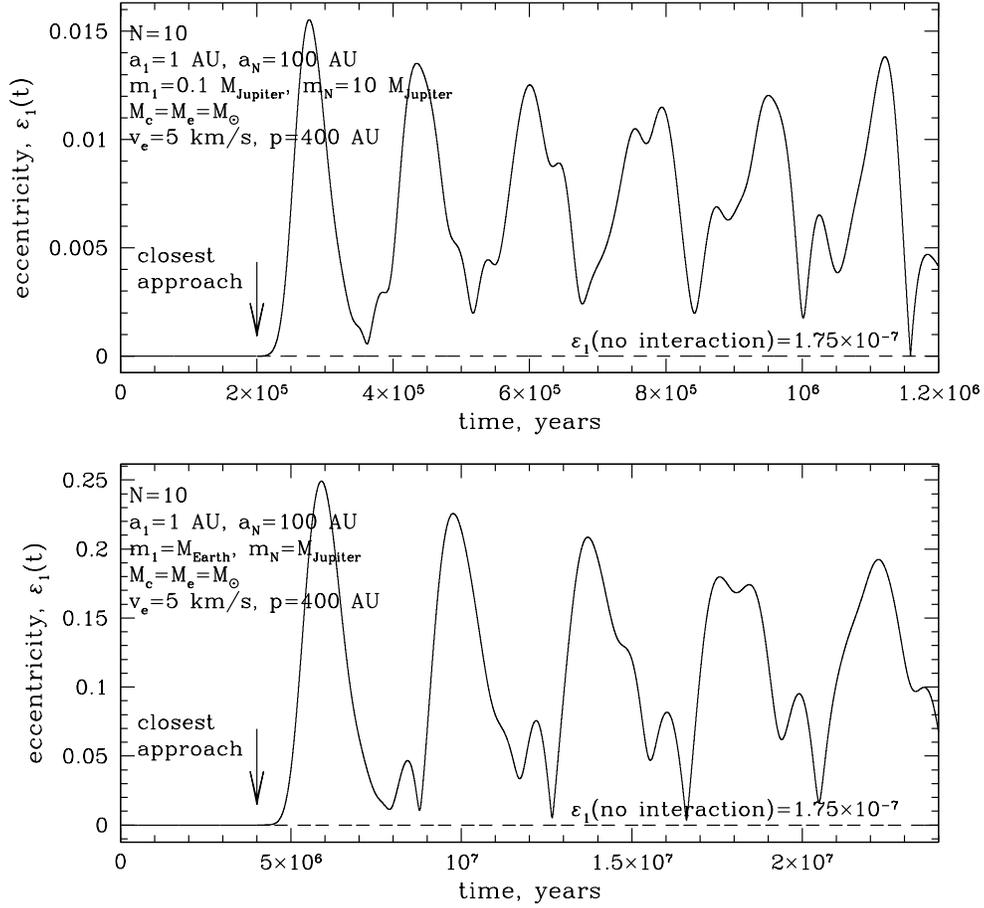}
\figcaption{Evolution of the eccentricity of the innermost planet due to secular interactions with other planets in two different systems. Top: masses of the planets range from 0.1 $M_{Jupiter}$ to 10 $M_{Jupiter}$. Bottom: masses of the planets range from $M_{Earth}$ to $M_{Jupiter}$. The parameters of the systems and of the perturbation are specified in the top left corner of each diagram. If there were no outer planets the eccentricity of the innermost planet would be just $1.75\times 10^{-7}$ as a result of the direct interaction with the perturber. In both cases the secular evolution is dominated by perturbations from the two outermost planets; the characteristic timescales are $P_{secular,i}\sim P_{orbital,i} M_c/m_i$ with $i=N$ producing the slow large amplitude variation and $i=N-1$ producing smaller amplitude variation.\label{pic_example}}
\end{figure}

The evolution of the eccentricity of the innermost planet is shown in Figure \ref{pic_example} (top). If there were no other planets in the system, a perturber passing with an impact parameter of 400 AU and with velocity $v_e=5$ km sec$^{-1}$ would only excite an eccentricity of $1.75\times 10^{-7}$. However, secular interactions with other planets bring the eccentricity up to a maximum of 0.015. The rms eccentricity can be found from equation (\ref{eccent_p}) to be $\sqrt{\left<\varepsilon_1^2(t)\right>}\simeq 0.008$. 

In a more extreme example given in Figure \ref{pic_example}, bottom, the spacing between the planets is the same but the masses increase from $m_1=M_{Earth}$ to $m_N=M_{Jupiter}$ with $g=1.90$. In this case secular interactions in the extended planetary system bring the eccentricity of the innermost planet up to 0.25, with an rms value of $\sqrt{\left<\varepsilon_1^2(t)\right>}\simeq 0.13$.

In both examples, the presence of outer planets made the perturbation much more efficient in exciting the eccentricity of the inner planet. We define for the inner planet an excitation efficiency parameter which is the ratio of the rms eccentricity after the perturbation to the value of eccentricity that would have been excited if there were no other planets:
\begin{equation}
{\mathbb X}\equiv \sqrt{\left<\varepsilon_1^2(\mbox{after perturbation})\right>}/\varepsilon_1(\mbox{excited if no interactions}).
\end{equation}
In the cases considered above, the excitation efficiency for the inner planet is ${\mathbb X}\simeq 4.6\times 10^4$ (Figure \ref{pic_example} top) and ${\mathbb X}\simeq 7.2\times 10^5$ (Figure \ref{pic_example} bottom).

Another important parameter is the ratio of the eccentricities of the innermost and the outermost planets. If this ratio is $\ll 1$, the eccentricity of the innermost planet cannot be excited significantly without ejecting the outer planet. We define a propagation efficiency parameter:
\begin{equation}
{\mathbb P}\equiv \sqrt{\left<\varepsilon_1^2(\mbox{after perturbation})\right>}/\sqrt{\left<\varepsilon_N^2(\mbox{after perturbation})\right>}.
\end{equation}
In the case of our example system in Figure \ref{pic_example} (top), the rms eccentricity of the outermost planet $\sqrt{\left<\varepsilon_N^2(t)\right>}\simeq 0.029 $, so the propagation efficiency is ${\mathbb P}\simeq 0.28$. In the second example (Figure \ref{pic_example} bottom), $\sqrt{\left<\varepsilon_N^2(t)\right>}\simeq 0.029$ and  ${\mathbb P}\simeq 4.4$. So in fact, the final rms eccentricity of the innermost planet is the largest in this system even though this planet is least affected by the direct interaction with the perturber.

Both efficiency parameters are quite high for our example systems, but they depend significantly on the mass distribution. The latter can be parametrized by the surface density that is obtained by smearing out the mass of all planets in such a way that the mass in the annulus between $a_i$ and $a_{i+1}$ is $m_i$. The resulting smeared-out surface density is a power law as a function of the distance from the central star, and the power index is related to the ratios of semimajor axes and masses:
\begin{equation}
\Sigma(r)\propto r^{-q}, \quad q=2-\ln(g)/\ln(f).
\end{equation}

We now proceed to investigate how the efficiencies ${\mathbb X}$ and ${\mathbb P}$ depend on the parameters of the planetary system. We calculated rms eccentricities for 16,000 simulated systems, with the total number of planets $N$ ranging from 5 to 20. The planets follow a geometric progression in semimajor axes and masses quantified by the parameters $f=a_{i+1}/a_i$ and $g=m_{i+1}/m_i$. All planets move in the same direction and are on initially circular orbits, and the perturber always has the same parameters ($M_e=M_{\odot}$, $p=400$ AU, $v_e=5$ km sec$^{-1}$). For each $N$, the parameter $f$ was selected randomly, with the maximum size of the system $a_N$ restricted to be in the range 50--200 AU. For each pair $N,f$ the parameter $g$ was selected randomly, but the ratio of the maximum mass to the minimum mass was not to exceed $5 M_{Jupiter}/M_{Earth}\simeq 1600$. For each set $N,f,g$ we selected random sets of phases $\varphi_i$ and calculated the resulting rms eccentricities for both prograde and retrograde motions of the perturber relative to the planets. The innermost planet was placed at 1 AU from the central star in all systems, so that in all systems
\begin{equation}
{\mathbb X}=\sqrt{\left<\varepsilon_1^2(t)\right>}/1.75\times 10^{-7}.
\end{equation}

\begin{figure}
\epsscale{0.8}
\plotone{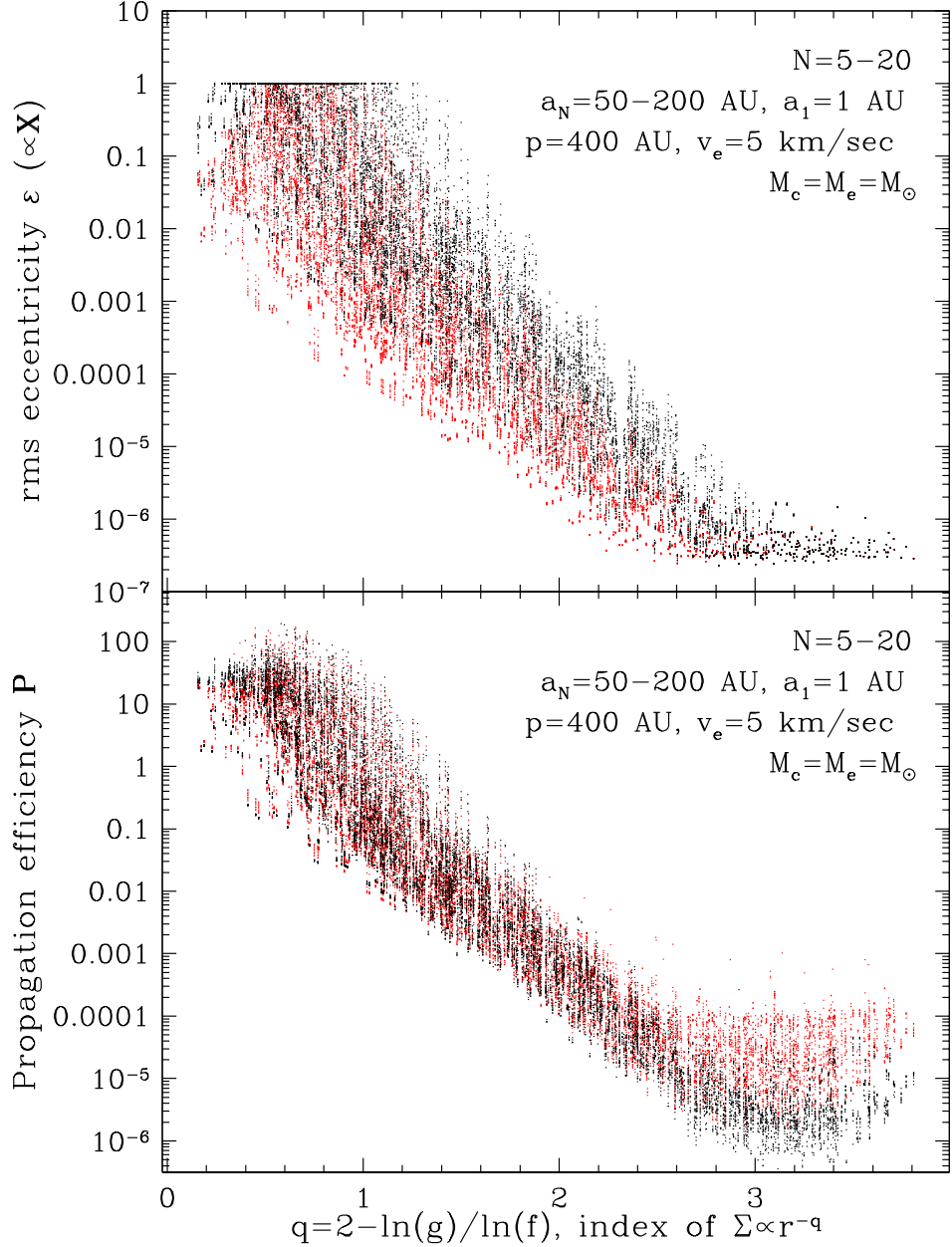}
\figcaption{Excitation and propagation efficiencies for a slow perturbation as a function of the smeared-out surface density distribution parameter. Prograde encounters are shown in black, and retrograde ones are shown in red; the parameters of the simulation are listed in the top right corner of both panels. In our ensemble of simulated systems all innermost planets are located at 1 AU and the parameters of the perturber are the same; therefore, the excitation efficiency is directly proportional to the rms eccentricity of the inner planet: ${\mathbb X}=\sqrt{\left<\varepsilon_1^2(t)\right>}/1.75\times 10^{-7}$. Systems in which the outer planet is stripped by the perturbation are not shown, and systems with very high calculated eccentricities $>$1 are shown as $\sqrt{\left<\varepsilon_1^2(t)\right>}=1$. \label{pic_stat}}  
\end{figure}

In Figure \ref{pic_stat} we plot the rms eccentricity of the innermost planet and the propagation efficiency for this planet as a function of the surface density index $q$. Most notably, both the excited eccentricity and the propagation efficiency are high in systems with flat mass distributions $q\la 1$. Prograde encounters are, on average, about ten times more efficient in exciting the eccentricity than retrograde encounters. It can be seen from Figure \ref{pic_stat} that high eccentricities can be excited for flat density distributions. The Laplace-Lagrange secular perturbation theory cannot be used for eccentricities approaching unity, and therefore we do not show systems with $\sqrt{\left<\varepsilon_1^2(t)\right>}\ga 1$ (numerical integrations showing the effects of secular interactions of high-eccentricity planets are discussed by \citealt{terq02}). We also do not show systems in which the outer planet is ejected as a result of the direct interaction with the perturber (this occurred in about 9\% of systems with sizes in excess of 100 AU; all planets with semimajor axes less than 100 AU remained bound to the central star). 

\begin{figure}
\epsscale{0.8}
\plotone{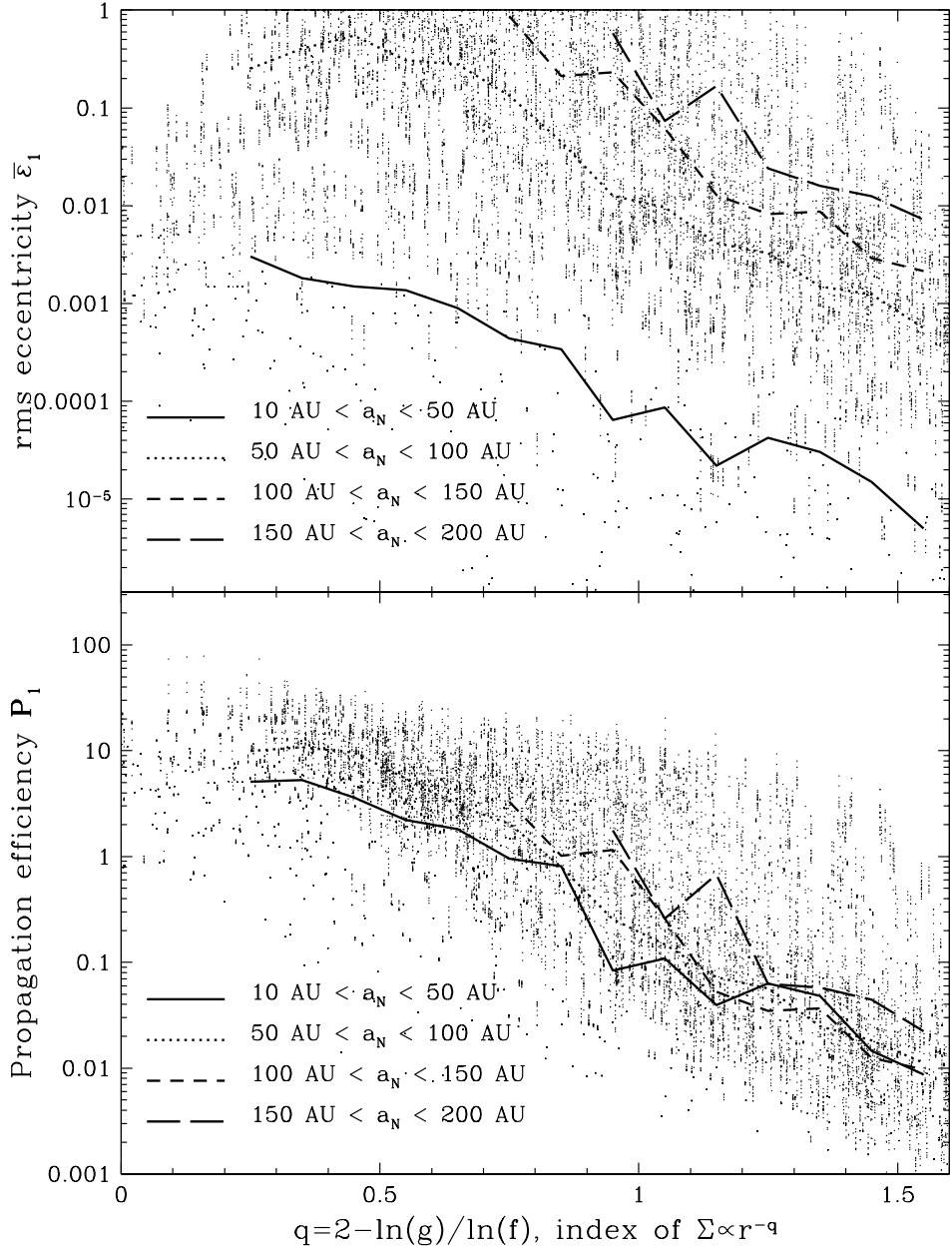}
\figcaption{Dependence of excitation and propagation efficiencies on the total size of the system. Only prograde encounters are shown, and we zoomed on the high-efficiency part of Figure \ref{pic_stat}. All systems are divided in four groups according to their size, and the median efficiency parameters are shown as a function of the mass distribution parameter $q$ for each of the three size groups. The total number of planets in this simulation varied between 5 and 20, the innermost planet was always at $a_1=1$ AU and the parameters of the impact were the same: $p=400$ AU, $v_e=5$ km sec$^{-1}$, $M_c=M_e=M_{\odot}$. The excited eccentricity is insensitive to the total size of the system $a_N$ as long as $50<a_N<200$, but drops significantly in systems smaller than 50 AU. \label{pic_size}}
\end{figure}

Our simulated systems occupy only a narrow locus on the $q$ vs efficiency diagrams, but this effect is artificial and is introduced by the restrictions we imposed on the simulated systems. In particular, we allowed only a narrow range of total sizes of the systems (50--200 AU). It is natural to assume that the excited eccentricities also depend on the total size of the system and, regardless of the mass distribution, high eccentricity cannot be excited in small systems. We therefore added to our simulation 11,000 more systems with sizes in the range 10 AU $<a_N<$ 50 AU and studied the dependence of the excited eccentricity on the total size of the system. For the encounter that we studied ($p=400$ AU, $v_e=5$ km sec$^{-1}$), the excited eccentricity is almost insensitive to size for 50 AU $<a_N<$ 200 AU, but drops significantly for smaller systems (Figure \ref{pic_size}).

Within the Laplace-Lagrange theory the final eccentricities do not depend on the overall mass scaling, but only on the ratios of the masses of the planets. The stability of the system, however, depends on the actual masses. \citet{glad93} gives a stability criterion for two-planet systems; if this criterion is applied to our simulations, the planetary system is stable if $f-1>3(m_{max}/M_c)^{1/3}$ where $m_{max}$ is the maximum mass of the planets in the system and $M_c$ is the mass of the host star. All simulations with $5\le N \le 20$ and the total sizes $50-200$ AU are stable by this criterion if $m_{max}<0.4 M_J$. If higher masses are allowed, the stability criterion limits the number of planets in the system. For example, if $m_{max}=5M_J$, then only systems with $N\le 13$ can be stable. The actual requirements for the long-term stability of multi-planet systems are likely to be somewhat more severe than Gladman's criterion suggests. Both excitation and propagation efficiencies are insensitive to the number of the planets in the system, and excluding unstable systems does not affect any results of our analysis. 

We also performed the same simulations for fast encounters ($p=400$ AU, $v_e=80$ km sec$^{-1}$). The excited eccentricities are typically about two magnitudes lower than in the case of slow encounters, in agreement with Figure \ref{pic_example}, but the propagation efficiencies are very similar to those shown in Figure \ref{pic_stat}. The main factors governing the efficiency of the eccentricity excitation and propagation are the total size of the planetary system and its large-scale surface density distribution, with efficient eccentricity excitation and propagation in systems with $q\la 1$. This conclusion is valid both for fast and slow encounters that we examined.  

\section{Discussion}
\label{sec_disc}

We showed that encounters with nearby stars can excite the eccentricities of the outer planets, which can then propagate to the inner planets. Eccentricity propagation is efficient for flat mass distributions of the planetary system. We now discuss the typical values of eccentricities excited by this mechanism in different environments. 

In the solar neighborhood, all encounters are fast by criterion (\ref{cond_fast}), and the impulse approximation works fairly well for planets at radii of a few tens of AU (Figure \ref{pic_excit} top). After the eccentricity excitation propagates to the innermost planet, its rms eccentricity becomes
\begin{equation}
{\rm rms}(\varepsilon_1)=0.0025 \mathbb P \left(\frac{v_e}{80 \mbox{ km sec}^{-1}}\right)^{-1} \left(\frac{a_N}{100 \mbox{ AU}}\right)^{3/2} \left(\frac{p}{500 \mbox{ AU}}\right)^{-2}=0.01-0.1 \label{eps_solar}
\end{equation}
for $\mathbb P=4-40$. We have averaged over all phases of closest approach in equation (\ref{fast}).

In an open cluster, the secular approximation (\ref{approx_rasio}) sets a lower bound on the excited eccentricity (Figure \ref{pic_excit} bottom), and for planets at radii larger than 50 AU the average excited eccentricities are about a factor of 3 larger than those given by the secular approximation (\ref{approx_rasio}). The rms eccentricity of the inner planet is then
\begin{equation}
{\rm rms}(\varepsilon_1)=0.053 \mathbb P \left(\frac{v_e}{5 \mbox{ km sec}^{-1}}\right)^{-1} \left(\frac{a_N}{100 \mbox{ AU}}\right)^{5/2} \left(\frac{p}{400 \mbox{ AU}}\right)^{-3}>0.2 \label{eps_open} 
\end{equation}
for $\mathbb P>4$. 

Typical eccentricities excited in dense open clusters are a factor of 20 greater than those that can be excited by field stars in the solar neighborhood. In the simulations shown in Figure \ref{pic_stat} propagation efficiencies of $\mathbb P=1-100$ were achieved for flat mass distributions ($q\la 1$). If such propagation efficiencies are common, the high eccentricities typical of those observed in extrasolar planets can be produced by close encounters in birth clusters during the $\sim 10^8$ years after the stars and planetary systems are formed. Future observations of systems containing extrasolar planets may determine whether there is enough mass in the outer parts of these systems for the suggested mechanism to be efficient. The observation times required to detect radial velocity signatures from companions at radii $>50$ AU are prohibitively long, but such objects can be detected directly by infrared imaging using the {\it Spitzer} telescope or by high-contrast imaging. The suggested mechanism also operates if the mass in the outer parts of the system is in the form of planetesimals. The detectability of such disks depend on their physical conditions.

The probability that the closest encounter during the life time of the system $\tau$ is $p$ or smaller is given by Poisson's formula:
\begin{equation}
\Pi(<p)=1-e^{-\tau n_* v_e \pi p^2}.\label{poisson}
\end{equation}
Here $n_*$ is the number density of the stars that act as perturbers. Therefore, about 10\% of all stars experience very close encounters (within 200 AU). In these cases the excited eccentricity can be very high even for field encounters in the solar neighborhood and is very sensitive to the size of the system and the phase of the outer planet at closest approach of the perturber. 

\section{Conclusions}
\label{sec_conc}

The eccentricities of most extrasolar planets are much larger than those of giant planets in the Solar System. This observation requires explanation because planets that form from a protoplanetary disk are expected to have nearly circular orbits. In this paper we explore the possibility that close stellar encounters can excite planetary eccentricities.

Encounters with nearby stars can be dynamically important for the outer parts of circumstellar disks and planetary systems. \citet{koba01} studied pumping up of eccentricities and inclinations of planetesimals in a close stellar encounter. \citet{kala01} and \citet{larw01} suggested that encounters with nearby stars can be responsible for asymmetries in the $\beta$ Pictoris disk. \citet{hurl02} found that most planets are stripped from their host stars by encounters with neighbor stars in the dense centers of globular clusters. 

In the solar neighborhood or an in open cluster the eccentricities of short-period planets cannot be excited directly by perturbations from nearby stars to any significant values because such planets are tightly bound to their host star. We suggest here that if there are other planets or a significant mass in smaller bodies in the same system on larger orbits, the outer planets can be excited to high eccentricities and then transfer this excitation to inner planets through secular perturbations. 

The extent to which the eccentricity of the innermost planets can be excited depends on the total size of the planetary system and on its smeared-out surface density distribution. If the smeared-out surface density distribution is parametrized by a power law, $\Sigma(r)\propto r^{-q}$, then the proposed mechanism is important for systems with $q\la 1$ and with total size $a_N\ga 50$ AU. At least some gas disks around young stars show flat density distributions with $q\simeq 1$ \citep{dutr96,wiln00} and extend out to tens or hundreds of AU. Theoretical models of steady-state protoplanetary disks also suggest flat surface density distributions \citep{bell97}. It appears that the maximum lifetime of such gas disks is $10^6-10^7$ years \citep{hais01}; some of this material may dissipate, but much or most of it may survive as planets or large planetesimals. Outward migration of planets can also yield massive planets at a few tens of AU \citep{vera04}. It has been suggested that the asymmetries in the dust debris disks in some systems may be due to unseen planets at radii $30-60$ AU \citep{ozer00, wiln02}. 

If the mass distribution is flat and the planetary system is large, the inner planets can be excited to high eccentricities (0.1$-$1, eq. \ref{eps_open}) due to encounters in rich birth clusters. The typical eccentricities that can be excited by the described mechanism from field star encounters in the solar neighborhood are smaller (0.01$-$0.1, eq. \ref{eps_solar}). A small fraction of stars (10\%) experience very close encounters with field stars (within 200 AU) in which excited eccentricities can be much higher.

All the calculations were done under a simplifying assumption that the perturber is moving in the same plane as the planets. \citet{koba01} argued that the eccentricities excited by a passing star are insensitive to the inclination of the perturber relative to the planetary disk and lie in the range determined by coplanar prograde and retrograde encounters. Therefore, our conclusions about the eccentricity excitation and propagation are likely to be independent of the inclination of the perturber. In the case when the trajectory of the perturber is inclined relative to the plane of the planetary system, inclinations of the outer planets in the system can be excited and then propagate inward, similarly to propagation of eccentricities that we discussed in this paper. Warps in circumstellar disks like those seen in $\beta$ Pictoris \citep{kala95, wahh03} may be signatures of inclination waves propagating through the system after the initial perturbation. 

In this work our focus was to find the conditions for efficient propagation of eccentricity disturbances. In our simulations, the masses and semimajor axes of successive planets follow geometric progressions. Clearly, these special configurations do not represent the full range of properties of real systems. Therefore, we do not discuss the distribution of eccentricities resulting from our simulations in relation to the observed distribution.

\bigskip
We would like to thank Marc Kuchner and Roman Rafikov for useful discussions and the referee for helpful comments on the manuscript. ST acknowledges the support of NASA grants NAG5-10456 and NNG04GH44G.

\newpage

\clearpage

\clearpage

\clearpage

\clearpage

\clearpage

\end{document}